\documentclass[fleqn,12pt,twoside]{article}
\usepackage{espcrc1}
\usepackage{epsf}

\usepackage{graphicx}
\usepackage[figuresright]{rotating}

\newcommand{\AmS}{{\protect\the\textfont2
  A\kern-.1667em\lower.5ex\hbox{M}\kern-.125emS}}

\title{Finite-size scaling for two-flavor QCD and
       comparison with $O(N)$ models}

\author{Tereza Mendes\address[MCSD]{IFSC--USP, Caixa Postal 369,
        13560-970 S\~ao Carlos SP, Brazil}}
       
\begin{document}

\maketitle

\begin{abstract}
The chiral transition for two-flavor QCD is predicted
to be in the same universality class as the
$3d$ $O(4)$ model. This prediction is verified 
in the Wilson case, but not for the staggered-fermion case.
The comparison is usually done assuming infinite-volume behavior.
Here we make an analysis of existing staggered-fermion data
using finite-size scaling and normalizing the QCD data. We find better
agreement for {\em larger} quark masses.
\end{abstract}

\vskip 6mm
In order to check if two systems belong to the
same universality class, one looks at universal
quantities, such as the critical exponents.
More generally, one compares scaling functions
for both systems.  The scaling ansatz
implies that the magnetization $M$ of a system be described by 
a universal function
\begin{equation} 
M/h^{1/\delta} = 
f_G(t/h^{1/\beta \delta})
\end{equation}
where $t\equiv (T-T_c)/T_0$ and $h\equiv H/H_0$ are the reduced 
temperature and magnetic field, respectively. 
Thus, once the non-universal normalization constants 
$T_0$ and $H_0$ are determined for a given system in the 
universality class, its order parameter M scales according to the
same scaling function $f_G$ for all systems in this class.
For comparisons with QCD, an important region is the
pseudo-critical line, defined by the points where
the susceptibility $\chi$ shows a (finite) peak. This
corresponds to the rounding of the divergence observed 
for $H=0, T=T_c$. The susceptibility scales as
$\chi \,=\, \partial M/\partial H \,=\,
(1/H_0)\,h^{1/\delta - 1}
\,g(t/h^{1/\beta\delta})$
where
$g(z) \,=\, 1/\delta\,[f_G(z) - z/\beta f_G'(z)]$.
At each fixed $h$ the peak in $\chi$ is given by
$t_{p}\,=\, z_p\,h^{1/\beta\delta}$
and we have 
$M_p\,=\, h^{1/\delta}\,f_G(z_p)$, 
$\,H_0\,\chi_{p}\,=\, h^{1/\delta - 1} \,
g(z_p)\,.$
Thus, the behavior along the pseudo-critical line is determined by the
universal constants $z_p$,$\,f_G(z_p)$,$g\,(z_p)$.

In addition to these infinite-volume scaling laws we may also 
consider finite-size-scaling functions. In fact, the scaling ansatz 
also implies
\begin{equation}
M = L^{-\beta/\nu} \,
Q_z(h\,L^{\beta\delta/\nu})
\end{equation}
where $L$ is the linear size of the system
and we consider fixed values of the ratio
$t/h^{1/\beta \delta}
\equiv z$
(e.g.\ $z=0$ as in the critical isotherm, 
or $z_p$ as along the pseudo-critical line).
Thus, $M$ can be described by a universal 
finite-size-scaling (FSS) function of one variable.
We note that in order to recover the infinite-volume expression
$M=h^{1/\delta}\,f_G(z)$
as $L\to\infty$, we must have
$\;Q_z(u)\,\to\,
f_G(z)\,u^{1/\delta}\;$
for large $u$. Thus, in this limit,
the FSS functions are given simply in terms of
the scaling function $f_G(z)$.
Working with the FSS functions $Q_z$ instead of
the infinite-volume scaling function $f_G$ has the
disadvantage that one must consider $z$ fixed
(thus restricting the regions to be compared in parameter space)
but the advantage that a comparison can be made
already at finite values of $L$.
 
\begin{figure}[htb]
\vspace{-2mm}
\begin{minipage}[t]{78mm}
\epsfxsize = \textwidth
\centerline{\leavevmode\epsffile{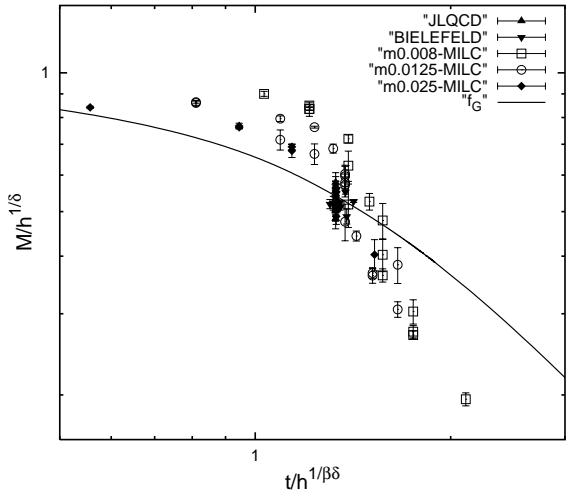}}
\vspace{-9.5mm}
\caption{Comparison at infinite volume.}
\label{1}
\end{minipage}
\hspace{\fill}
\begin{minipage}[t]{78mm}
\epsfxsize = \textwidth
\centerline{\leavevmode\epsffile{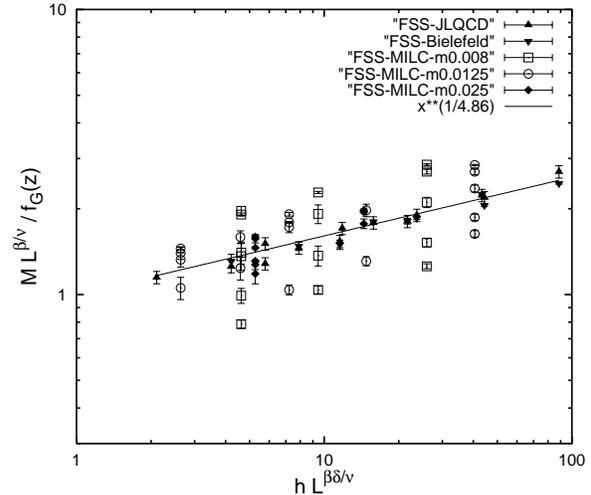}}
\vspace{-9.5mm}
\caption{same, using finite-size scaling.}
\label{2}
\end{minipage}
\vspace{-7.5mm}
\end{figure}

Critical exponents and the scaling function $f_G$
are well-known for the $3d$ $O(4)$ model, whereas
the pseudo-critical line and
FSS functions have been recently studied
in [1]. We find that the peak susceptibilities are given by
$z_p = 1.33(5)$
and that the FSS functions $Q_z(u)$ approach
their asymptotic values already at $u\sim 2$.
For QCD, the order parameter is given by the chiral condensate
$<\overline \psi \,\psi>$, the 
analogue of the magnetic field is the 
quark mass $m_q$, and $t$
is proportional to $6/g^2 - 6/g_c^2(0)$.
Thus, the chiral susceptibility peaks at 
$t_{p}\sim m_q^{1/\beta\delta}$.
Studies at $N_{\tau}=4$ by
the Bielefeld, JLQCD and MILC groups [2]
show that the peak positions scale
with the predicted exponents, but 
one sees no agreement with the $O(N)$ 
scaling function. This comparison was done up to
multiplicative constants in the fields $t$
and $h$. We redo it below,
performing a normalization of the QCD data.

Let $h\equiv N_{\tau}\,m_q/H_0$,
$t\equiv (6/g^2 - 6/g_c^2(0))/T_0$. From the
observed scaling along the pseudo-critical line
and the universal quantities 
$z_p$,$\,f_G(z_p)$ from the $O(4)$ model,
we determine the normalization constants
$H_0$ and $T_0$.
This allows an unambiguous 
comparison of the QCD data to $f_G$,
as shown in Fig.\ \ref{1}.
We then do a comparison with the FSS functions
$Q_z(h\,L^{\beta\delta/\nu})$ using
the data normalized above. We see that
along the pseudo-critical line (JLQCD and Bielefeld data)
FSS works well: the points are already in the asymptotic 
region and scale with the predicted exponent 
$\delta$ --- as shown in [1] --- and with the correct 
multiplicative constant, as expected from our normalization.
In fact, all QCD data are asymptotic in 
$h\,L^{\beta\delta/\nu}$,
and we can thus rescale all data by their corresponding
$f_G(z)$ factors and plot them together, as shown in Fig.\ \ref{2}.
We see a better agreement overall, although most MILC points
are several deviations away from the predicted curve.
As in Fig.\ \ref{1}, this happens for points
outside the pseudo-critical line. Among these points, the ones
with {\em larger} quark masses come closer to the curve.
In particular, for the points with $m_q=0.025$
we see reasonable agreement, suggesting that after finite-size effects
have been taken into account one may see universal scaling
even away from the pseudo-critical line.

\vskip 2mm
It is a pleasure to thank J\"urgen Engels and Philippe de Forcrand 
for helpful comments and suggestions. This work was supported by
FAPESP, Brazil (project No.\ 00/05047-5).

\vskip 2mm
\noindent
[1] J. Engels, S. Holtmann, T. Mendes and T. Schulze,
                  Phys.\ Lett.\ B514, 299 (2001).

\noindent
[2] F. Karsch and E. Laermann, Phys.\ Rev.\ D50, 6954 (1994);
            S. Aoki et al., Phys.\ Rev.\ 
            \phantom{all} D57, 3910 (1998);
            C. Bernard et al., Phys.\ Rev.\ D61, 054503 (2000).

\end{document}